\documentclass[conference,10pt]{IEEEtran}
\usepackage{url}
\usepackage[utf8]{inputenc}
\usepackage{xcolor}
\usepackage{amsmath}
\usepackage{amssymb}
\usepackage{array}

\usepackage{geometry}
\usepackage[acronyms,nonumberlist,nopostdot,nomain,nogroupskip]{glossaries}
\usepackage{tablefootnote}
\usepackage{booktabs}

\usepackage{tabularx}
\usepackage{physics}
\usepackage{siunitx}
\usepackage{pgfplots}
\usepackage[outdir=img/]{epstopdf}
\pgfplotsset{compat=newest} 
\pgfplotsset{plot coordinates/math parser=false} 
\newlength\fheight
\newlength\fwidth
\usetikzlibrary{plotmarks,patterns,decorations.pathreplacing,backgrounds,calc,arrows,arrows.meta,spy,matrix}
\usepgfplotslibrary{patchplots,groupplots}

\def\BibTeX{{\rm B\kern-.05em{\sc i\kern-.025em b}\kern-.08em T\kern-.1667em\lower.7ex\hbox{E}\kern-.125emX}}

\usepackage{enumitem}
\usepackage[capitalize]{cleveref}
\usepackage{caption}
\usepackage{multirow}
\usepackage{hhline}

\usepackage[font=scriptsize]{subcaption}
\usepackage[font=footnotesize]{caption}

\usepackage{mathtools}

\newacronym{3gpp}{3GPP}{3rd Generation Partnership Project}
\newacronym{5g}{5G}{5th generation}
\newacronym{5gc}{5GC}{5G Core}
\newacronym{adc}{ADC}{Analog to Digital Converter}
\newacronym{aimd}{AIMD}{Additive Increase Multiplicative Decrease}
\newacronym{am}{AM}{Acknowledged Mode}
\newacronym{amc}{AMC}{Adaptive Modulation and Coding}
\newacronym{aqm}{AQM}{Active Queue Management}
\newacronym{awgn}{AGWN}{Additive White Gaussian Noise}
\newacronym{balia}{BALIA}{Balanced Link Adaptation}
\newacronym{bdp}{BDP}{Bandwidth-Delay Product}
\newacronym{bf}{BF}{Beamforming}
\newacronym{bs}{BS}{Base Station}
\newacronym{cc}{CC}{Congestion Control}
\newacronym{cdf}{CDF}{Cumulative Distribution Function}
\newacronym{cn}{CN}{Core Network}
\newacronym{cp}{CP}{Control Plane}
\newacronym{cqi}{CQI}{Channel Quality Information}
\newacronym{crs}{CRS}{Cell Reference Signal}
\newacronym{csirs}{CSI-RS}{Channel State Information - Reference Signal}
\newacronym{dc}{DC}{Dual Connectivity}
\newacronym{dce}{DCE}{Direct Code Execution}
\newacronym{dci}{DCI}{Downlink Control Information}
\newacronym{dl}{DL}{Deep Learning}
\newacronym{dmr}{DMR}{Deadline Miss Ratio}
\newacronym{dmrs}{DMRS}{DeModulation Reference Signal}
\newacronym{e2e}{E2E}{End-to-End}
\newacronym{ecn}{ECN}{Explicit Congestion Notification}
\newacronym{edf}{EDF}{Earliest Deadline First}
\newacronym{enb}{eNB}{evolved Node Base}
\newacronym{endc}{EN-DC}{E-UTRAN-\gls{nr} \gls{dc}}
\newacronym{epc}{EPC}{Evolved Packet Core}
\newacronym{es}{ES}{Edge Server}
\newacronym{fdd}{FDD}{Frequency Division Duplexing}
\newacronym{fdma}{FDMA}{Frequency Division Multiple Access}
\newacronym{fs}{FS}{Fast Switching}
\newacronym{ftp}{FTP}{File Transfer Protocol}
\newacronym{gnb}{gNB}{Next Generation Node Base}
\newacronym{gpr}{GPR}{Gaussian Process Regression}
\newacronym{harq}{HARQ}{Hybrid Automatic Repeat reQuest}
\newacronym{hpbw}{HPBW}{Half Power BeamWidth}
\newacronym{hetnet}{HetNet}{Heterogeneous Network}
\newacronym{hh}{HH}{Hard Handover}
\newacronym{hol}{HOL}{Head-of-Line}
\newacronym{hope}{HOpE}{High-dimensional OPtimization through Emulation}
\newacronym{hqf}{HQF}{Highest-quality-first}
\newacronym{ia}{IA}{Initial Access}
\newacronym{iab}{IAB}{Integrated Access and Backhaul}
\newacronym{imt}{IMT}{International Mobile Telecommunication}
\newacronym{inr}{INR}{Interference to Noise Ratio}
\newacronym{iot}{IoT}{Internet of Things}
\newacronym{kpi}{KPI}{Key Performance Indicator}
\newacronym{los}{LoS}{Line-of-Sight}
\newacronym{lte}{LTE}{Long Term Evolution}
\newacronym{m2m}{M2M}{Machine to Machine}
\newacronym{mac}{MAC}{Medium Access Control}
\newacronym{mc}{MC}{Multi-Connectivity}
\newacronym{mcs}{MCS}{Modulation and Coding Scheme}
\newacronym{mec}{MEC}{Mobile Edge Cloud}
\newacronym{mi}{MI}{Mutual Information}
\newacronym{mib}{MIB}{Master Information Block}
\newacronym{mimo}{MIMO}{Multiple-Input Multiple-Output}
\newacronym{ml}{ML}{Machine Learning}
\newacronym{mlr}{MLR}{Maximum-local-rate}
\newacronym[plural=\gls{mme}s,firstplural=Mobility Management Entities (MMEs)]{mme}{MME}{Mobility Management Entity}
\newacronym{mmwave}{mmWave}{millimeter wave}
\newacronym{mptcp}{MPTCP}{Multipath TCP}
\newacronym{mr}{MR}{Maximum Rate}
\newacronym{mrdc}{MR-DC}{Multi \gls{rat} \gls{dc}}
\newacronym{mss}{MSS}{Maximum Segment Size}
\newacronym{mtd}{MTD}{Machine-Type Device}
\newacronym{mtu}{MTU}{Maximum Transmission Unit}
\newacronym{nfv}{NFV}{Network Function Virtualization}
\newacronym{nlos}{NLoS}{Non-\gls{los}}
\newacronym{nr}{NR}{New Radio}
\newacronym{nrmse}{nRMSE}{normalized Root Mean Square Error}
\newacronym{nn}{NN}{Neural Network}
\newacronym{nyu}{NYU}{New York University}
\newacronym{nsa}{NSA}{Non Stand Alone}
\newacronym{o2i}{O2I}{Outdoor-to-Indoor}
\newacronym{ofdm}{OFDM}{Orthogonal Frequency Division Multiplexing}
\newacronym{pa}{PA}{Position-aware}
\newacronym{pbch}{PBCH}{Physical Broadcast Channel}
\newacronym{pdcch}{PDCCH}{Physical Downlink Control Channel}
\newacronym{pdcp}{PDCP}{Packet Data Convergence Protocol}
\newacronym{pdsch}{PDSCH}{Physical Downlink Shared Channel}
\newacronym{pdu}{PDU}{Packet Data Unit}
\newacronym{pf}{PF}{Proportional Fair}
\newacronym{pgw}{PGW}{Packet Gateway}
\newacronym{phy}{PHY}{Physical}
\newacronym{ppp}{PPP}{Poisson Point Process}
\newacronym{prb}{PRB}{Physical Resource Block}
\newacronym{pss}{PSS}{Primary Synchronization Signal}
\newacronym{pucch}{PUCCH}{Physical Uplink Control Channel}
\newacronym{pusch}{PUSCH}{Physical Uplink Shared Channel}
\newacronym{rach}{RACH}{Random Access Channel}
\newacronym{ran}{RAN}{Radio Access Network}
\newacronym[firstplural=Radio Access Technologies (RATs)]{rat}{RAT}{Radio Access Technology}
\newacronym{red}{RED}{Random Early Detection}
\newacronym{rf}{RF}{Radio Frequency}
\newacronym{rlc}{RLC}{Radio Link Control}
\newacronym{rlf}{RLF}{Radio Link Failure}
\newacronym{rmse}{RMSE}{Root Mean Square Error}
\newacronym{rr}{RR}{Round Robin}
\newacronym{rrc}{RRC}{Radio Resource Control}
\newacronym{rrm}{RRM}{Radio Resource Management}
\newacronym{rs}{RS}{Remote Server}
\newacronym{rsrp}{RSRP}{Reference Signal Received Power}
\newacronym{rsrq}{RSRQ}{Reference Signal Received Quality}
\newacronym{rss}{RSS}{Received Signal Strength}
\newacronym{rssi}{RSSI}{Received Signal Strength Indicator}
\newacronym{rtt}{RTT}{Round Trip Time}
\newacronym{rw}{RW}{Receive Window}
\newacronym{rx}{RX}{Receiver}
\newacronym{sa}{SA}{standalone}
\newacronym{sack}{SACK}{Selective Acknowledgment}
\newacronym{sap}{SAP}{Service Access Point}
\newacronym{sch}{SCH}{Secondary Cell Handover}
\newacronym{scm}{SCM}{Spatial Channel Model}
\newacronym{scoot}{SCOOT}{Split Cycle Offset Optimization Technique}
\newacronym{sdma}{SDMA}{Spatial Division Multiple Access}
\newacronym{si}{SI}{Study Item}
\newacronym{sib}{SIB}{Secondary Information Block}
\newacronym{sinr}{SINR}{Signal to Interference plus Noise Ratio}
\newacronym{sm}{SM}{Saturation Mode}
\newacronym{snr}{SNR}{Signal to Noise Ratio}
\newacronym{son}{SON}{Self-Organizing Network}
\newacronym{srs}{SRS}{Sounding Reference Signal}
\newacronym{ss}{SS}{Synchronization Signal}
\newacronym{sss}{SSS}{Secondary Synchronization Signal}
\newacronym{svm}{SVM}{Support Vector Machine}
\newacronym{svr}{SVR}{Support Vector Regressor}
\newacronym{tb}{TB}{Transport Block}
\newacronym{tcp}{TCP}{Transmission Control Protocol}
\newacronym{tdd}{TDD}{Time Division Duplexing}
\newacronym{tdma}{TDMA}{Time Division Multiple Access}
\newacronym{tfl}{TfL}{Transport for London}
\newacronym{tm}{TM}{Transparent Mode}
\newacronym{trp}{TRP}{Transmitter Receiver Pair}
\newacronym{tti}{TTI}{Transmission Time Interval}
\newacronym{ttt}{TTT}{Time-to-Trigger}
\newacronym{tx}{TX}{Transmitter}
\newacronym{ue}{UE}{User Equipment}
\newacronym{ul}{UL}{Uplink}
\newacronym{ula}{ULA}{Uniform Linear Array}
\newacronym{um}{UM}{Unacknowledged Mode}
\newacronym{umi}{UMi}{Urban Micro-cell}
\newacronym{uml}{UML}{Unified Modeling Language}
\newacronym{upa}{UPA}{Uniform Planar Array}
\newacronym{utc}{UTC}{Urban Traffic Control}
\newacronym{ut}{UT}{User Terminal}
\newacronym{vm}{VM}{Virtual Machine}
\newacronym{vr}{VR}{Virtual Reality}
\newacronym{wbf}{WBF}{Wired Bias Function}
\newacronym{wf}{WF}{Wired-first}

\newcommand{\sinr}{\ensuremath{\mathrm{SINR}}}
\newcommand{\sinrmean}{\ensuremath{\overline{\sinr}}}

\definecolor{desireRed}{RGB}{230,57,60}%
\definecolor{darkPurple}{RGB}{59,31,43}%
\definecolor{springGreen}{RGB}{37,223,145}%
\definecolor{queenBlue}{RGB}{69,123,157}%
\definecolor{spaceCadet}{RGB}{29,53,87}%

\makeglossaries

\begin{document}
  
\setlength{\extrarowheight}{2pt}
\glsunset{nr}
  
\title{Enabling Simulation-Based Optimization Through\\
  Machine~Learning:~A~Case~Study~on~Antenna~Design
  \vspace{-4ex}}

\author{{  \parbox{\linewidth}{ \centering
\textbf{\IEEEauthorrefmark{1}Paolo Testolina},
\textbf{\IEEEauthorrefmark{1}Mattia Lecci},
\textbf{\IEEEauthorrefmark{1}Mattia Rebato},
\textbf{\IEEEauthorrefmark{1}Alberto Testolin},\\
\textbf{\IEEEauthorrefmark{2}Jonathan Gambini},
\textbf{\IEEEauthorrefmark{2}Roberto Flamini},
\textbf{\IEEEauthorrefmark{2}Christian Mazzucco},
\textbf{\IEEEauthorrefmark{1}Michele Zorzi}}}\\
 \IEEEauthorrefmark{1}Department of Information Engineering, University of Padova, Italy \\
 \IEEEauthorrefmark{2}HUAWEI Technologies, Milan, Italy \\
$\{$\texttt{testolin, leccimat, rebatoma, zorzi}$\}$\texttt{@dei.unipd.it}; \texttt{alberto.testolin@unipd.it};\\
{$\{$\texttt{jonathan.gambini, roberto.flamini1, christian.mazzucco}$\}$ \texttt{@huawei.com}}
\vspace{-1ex}}

\makeatletter



\maketitle

\begin{abstract}
Complex phenomena are generally modeled with sophisticated simulators that, depending on their accuracy, can be very demanding in terms of computational resources and simulation time.
Their time-consuming nature, together with a typically vast parameter space to be explored, make simulation-based optimization often infeasible.
In this work, we present a method that enables the optimization of complex systems through \gls{ml} techniques.
We show how well-known learning algorithms are able to reliably emulate a complex simulator with a modest dataset obtained from it.
The trained emulator is then able to yield values close to the simulated ones in virtually no time.
Therefore, it is possible to perform a global numerical optimization over the vast multi-dimensional parameter space, in a fraction of the time that would be required by a simple brute-force search.
As a testbed for the proposed methodology, we used a network simulator for next-generation mmWave cellular systems.
After simulating several antenna configurations and collecting the resulting network-level statistics, we feed it into our framework.
Results show that, even with few data points, extrapolating a continuous model makes it possible to estimate the global optimum configuration almost instantaneously. 
The very same tool can then be used to achieve any further optimization goal on the same input parameters in negligible time.
\end{abstract}

\glsresetall

\begin{picture}(0,0)(-16,-295)
\centering
\put(0,0){
\put(-25,170){P. Testolina, M. Lecci, M. Rebato, A. Testolin, J. Gambini, C. Mazzucco and M. Zorzi, ``Enabling Simulation-Based}
\put(-25,160){Optimization Through Machine Learning: A Case Study on Antenna Design", in IEEE Global Communication}
\put(-25,150){Conference: Wireless Communication (GLOBECOM2019 WC), Waikoloa, USA, Dec 2019.}}
\end{picture}

\section{Introduction}
\label{sec:introduction}
Large antenna arrays and \gls{mmwave} frequencies will be adopted to meet the challenging requirements of future 5G mobile networks.
Due to the increased path loss at such high frequencies, multi-antenna systems with beamforming techniques are used to increase the antenna gain, thus making it possible to reach an acceptable communication range in \gls{mmwave} networks~\cite{rangan14}.

Among the possible antenna array designs, the most suitable approach is the use of \glspl{upa}, where the antenna elements are spaced on a two-dimensional plane and a 3D beam can be synthesized~\cite{Rabinovich}.
However, before proceeding with the manufacturing, a careful design phase is required to optimize the user performance.
For an accurate analysis of 5G \gls{mmwave} cellular scenarios, it is important to consider realistic antenna patterns combined with a rigorous channel model in order to simulate the wireless radiation environment \cite{rebato18}.

Given the high prototyping cost, antenna designs are generally evaluated through simulators first.
However, given the large number of parameters that need to be tuned, the optimization of an objective function (e.g., maximization of the \gls{sinr}) is extremely time-consuming, or even infeasible.
Gathering results from a realistic simulator can take a very long time, depending on the required level of detail and the accuracy of the employed antenna models.
Therefore, an alternative way must be found to reach the optimization goal.
\glsunset{ml}
\begin{figure}[t]
  \centering
  \includegraphics[width=1\columnwidth]{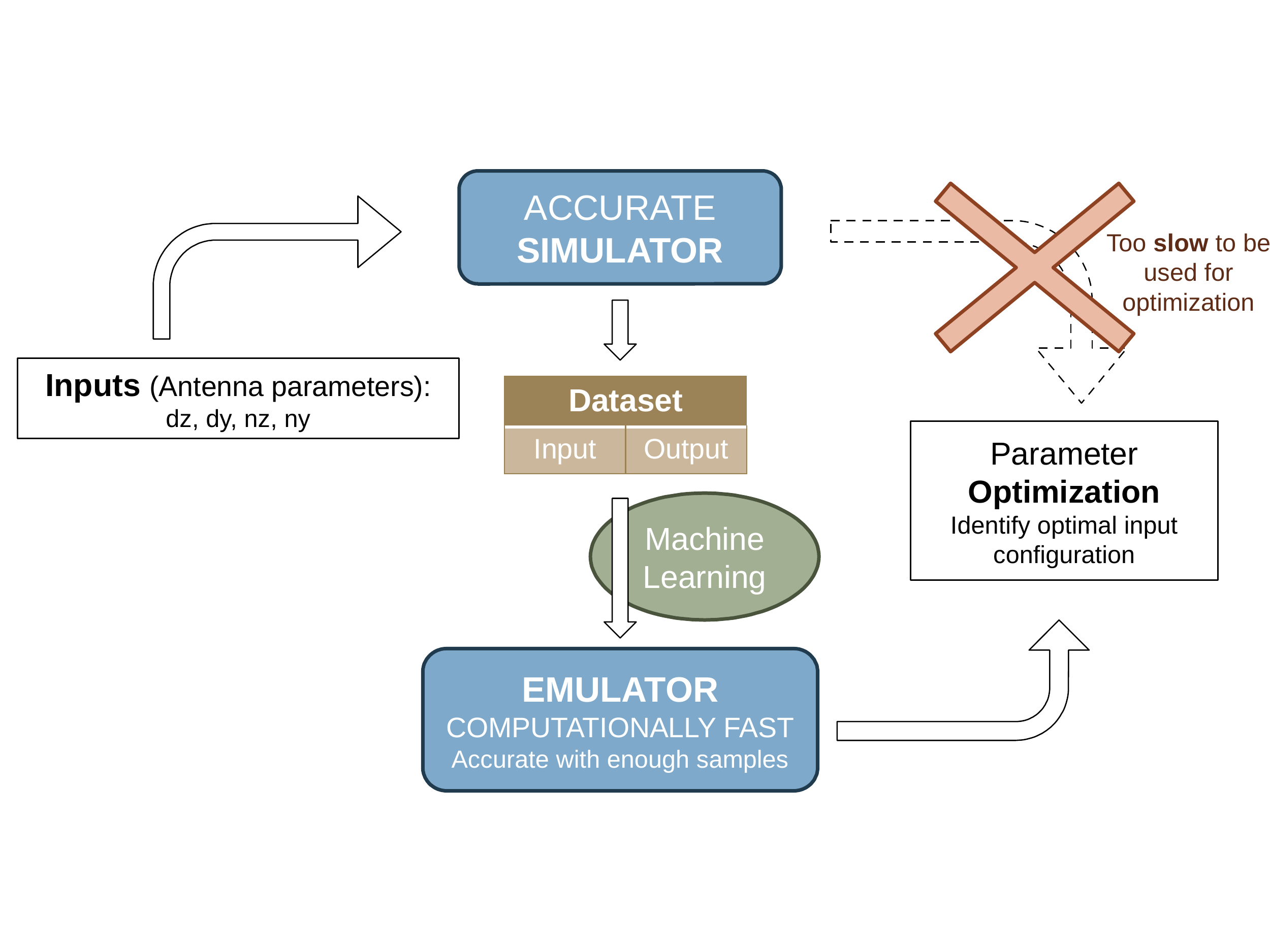}
  \caption{Workflow of the proposed framework.
  The diagram highlights how the \emph{parameter optimization} is reached using a ML-based emulator.}
  \label{fig:workflow}
\end{figure}

\glsreset{ml}
To address this problem, in this paper we propose and evaluate a \gls{ml} framework able to emulate a complex simulator and to achieve the optimization in a reasonable amount of time.
This concept is represented by the diagram in \cref{fig:workflow}, which shows how the parameter optimization can be reached through the \gls{ml}-based emulator, that only requires a relatively small dataset of simulated data.

To summarize, given the aforementioned objectives, in this paper we introduce a methodology able to greatly reduce the amount of time required to optimize the parameters of complex simulations by means of \gls{ml} algorithms.
We test and evaluate the proposed framework on a 5G network simulator, with the intention to optimize basic antenna array parameters.

\subsection{Related Works}
\label{subsec:related_works}
Researchers are eager to understand the possibilities that the \gls{ml} techniques can offer when applied to communication problems. The new database proposed in~\cite{Alkhateeb2019} is proof of this new trend, as it lays the premises for a common research ground. 

One common application of \gls{ml} is parameter estimation, where great results were achieved even where the most sophisticated classical techniques failed.

This is the case in~\cite{Arnold2019}, where the authors try to estimate the downlink channel starting from samples of the uplink channel.
While well-known signal processing techniques (e.g., the Wiener filter) were not able to perform a good estimate, the \gls{ml} approach proposed by the authors yields very good results.

Another common approach is the encoding of the channel representation through autoencoders~\cite{simeone}.
Autoencoders are an unsupervised learning algorithm, and as such they do not need labeled data but can learn autonomously.
The idea behind this technique is to train two \glspl{nn}, one performing the encoding of the input data, the second trying to decode it.
The layer between the two should contain, in our case, a useful and extremely compressed representation of the channel.
This can be applied at many levels, starting from the pure, physical channel model, to the entire transmitter-channel-receiver chain~\cite{OShea2017}.
This allows obtaining either encoders/decoders, transmitter/receiver chains or channel models that have a much lower computational complexity.

Then, \gls{ml} has been successfully applied also at the network layer.
Innovative ideas and proposals have challenged even the most resilient classical paradigms such as the ISO/OSI architecture~\cite{cognetworks}, as the evolution of the information infrastructures calls for a radical change.
These new approaches started showing their potential in the increasingly heterogeneous network scenarios, e.g., when facing the high data load and quality of experience required for video streaming~\cite{testolin}.
We believe that \gls{ml} can play a central role in tackling highly complex problems in a data-rich environment.

Furthermore, the authors in~\cite{sun18} use Deep \glspl{nn} to optimize the allocation algorithm in a wireless resource management problem.
The proposed concept is similar to the one described in our work, as a learning tool is used to approximate a complex input-output function.
However, the authors also include the optimization step into the learning process and use many more training samples to accommodate the needs of their deep architecture.
For our work, instead, it is crucial to use as few samples as possible as we aim to speed up the optimization process by approximating very slow simulators, making the data acquisition the main bottleneck.

In the literature, many research activities have been focusing on the study of \gls{mmwave} mobile environments while in parallel a lot of works have studied in the past the problem of beamforming and antenna array optimization.
However, there are no conclusive works focused on antenna optimization for \gls{mmwave} mobile scenarios.

In view of this goal, in the remainder of the section, we report some related works on antenna characterization for \gls{mmwave} bands which have been a guideline for the activity carried out in this paper.
Our previous work~\cite{Rebato16} has been considered as the baseline for the mobile network simulator.
Starting from that, several changes have been made to adapt the cellular simulator to test all the antenna element and array settings.

At high frequencies, such as in the \gls{mmwave} bands, where strong attenuations are present, quantifying the actual antenna gain obtained due to the radiation pattern is fundamental to precisely evaluate any \gls{mmwave} system.
In a previous work~\cite{rebato18}, realistic antenna radiation patterns have been studied and precisely characterized, motivated by the need to properly capture \gls{mmwave} propagation behaviors and understand the achievable performance in 5G cellular scenarios.
As it is customary, antenna patterns were modeled as the superposition between the single element radiation pattern and the array factor.
The latter term is a function used to characterize the effects of the entire array, while the former is used to quantify how the power of each antenna element is irradiated in all directions.
The work shows how the single element radiation pattern affects the network performance, thus highlighting how optimization of this further parameter is not only possible but also required.

\section{Framework Description}
\label{sec:framework_description}

\begin{figure*}[t]
    \centering
    \includegraphics[width=1\textwidth]{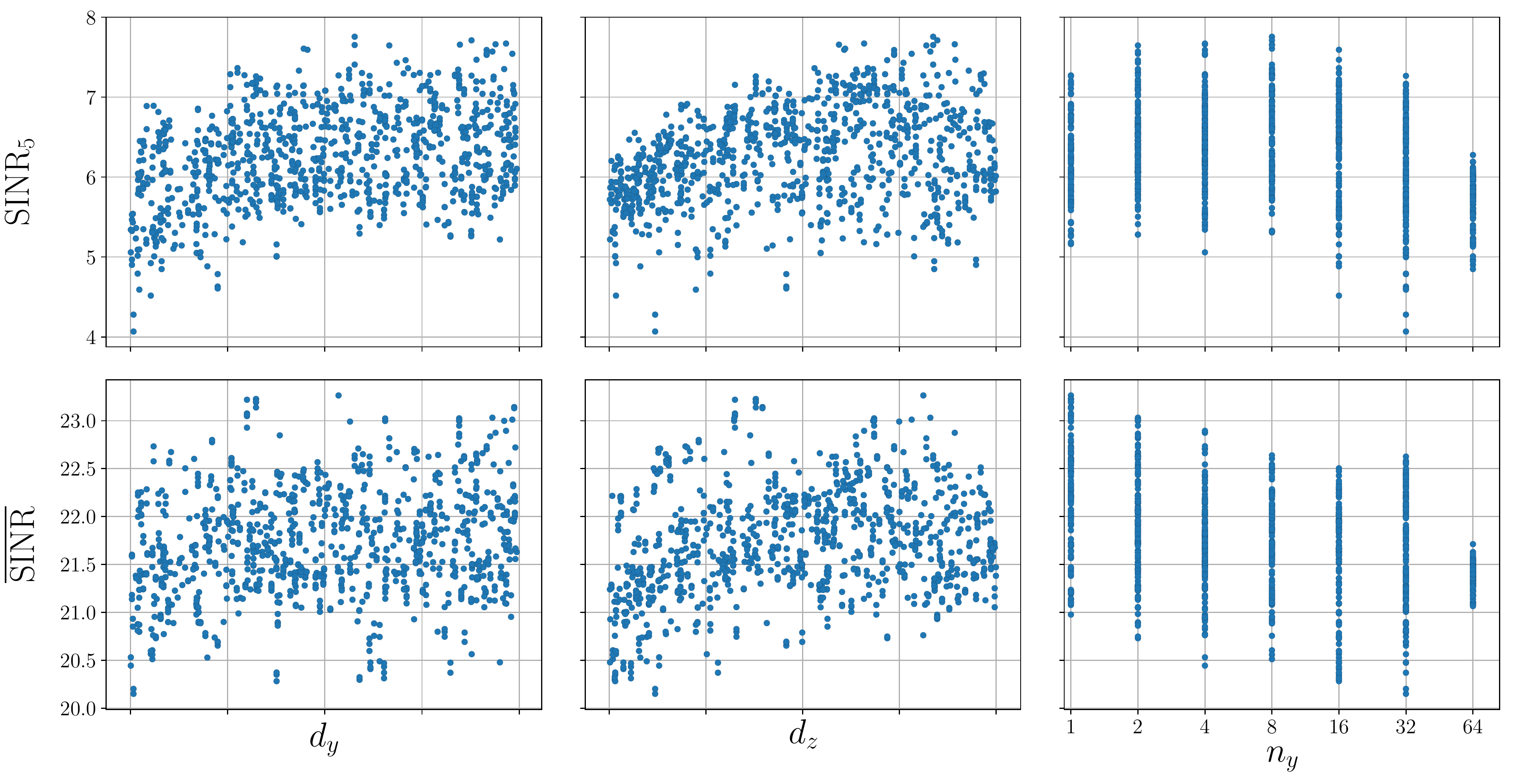}
    \caption{Correlation between selected inputs and outputs.}
    \label{fig:pairplot}
\end{figure*}

The objective of the proposed framework is to speed up simulation-based optimization in the presence of slow simulators.
Optimization based on simulated data requires several iterations, each with a different input configuration, for the optimization strategy to steer toward the optimal value.
The major constraint is the simulation time\footnote{
Simulation times vary remarkably depending on the type of simulation and the accuracy required.
It is not unlikely for a single run to require hours or even days.
}, which makes a brute-force approach infeasible.
The goal of our framework is to require a small number of simulations to learn the input-output relationship through \gls{ml} algorithms, which are orders of magnitude faster to evaluate.
A key advantage is that, after the preliminary database creation, the optimization of the selected antenna parameters can be achieved in a negligible amount of time, even when testing different optimization goals.
In fact, we remark that once the emulator is trained, the optimization of multiple objective functions can be done instantaneously.

Although the idea is broadly applicable, our focus here is on antenna optimization over network-level metrics for \gls{mmwave} systems.
We consider this use case as a testbed and we report the results of this particular optimization later in the paper.

In order to assess the validity of this framework, three main questions have to be answered:
\begin{enumerate}[label=Q.\arabic*,ref=Q.\arabic*]
  \item \label{item:1}Is it possible to emulate a complex network simulator with a learning tool, and which learning tool can achieve the best emulation accuracy?
  \item \label{item:nSamples}How many train and test samples are needed for the chosen learning paradigm to converge and to be robustly evaluated?
  \item Does the achieved precision allow an optimization that is accurate enough to be useful?
\end{enumerate}
The remainder of this section is devoted to addressing these problems.

\subsection{Network Simulator}
\label{sec:sim}
To test the framework, we need some data to learn from.
A custom simulator was built in order to efficiently obtain results from such complex simulations.
Simulation parameters are \gls{3gpp} standard compliant~\cite{3gpp.38.901,3gpp.38.913}, using the \gls{umi} scenario with no \gls{o2i} losses.

The variable parameters of the scenarios are listed here:
\begin{itemize}
  \item the antenna spacing $d_z, d_y $ in the vertical and horizontal directions, respectively;
  \item the number of antenna elements $n_z, n_y \in \{1,2,\ldots,N\}$ in the vertical and horizontal directions, respectively.
  The total number of antenna elements is fixed to $N = 64$, in order to obtain a fair comparison between different configurations.
  Thus, $n_z$ and $n_y$ are the integer divisors of $N$ and are deterministically related through $n_z = N/n_y$.
\end{itemize}
For each configuration, we collect network-level metrics such as:
\begin{itemize}
\item the average \gls{sinr} $\qty(\sinrmean)$;
\item the $5^{th}$ percentile of the \gls{sinr} $\qty(\sinr_5)$.
\end{itemize}

\subsection{Data Analysis and Machine Learning}
\label{sec:ml}
\glsreset{nn}
The dataset was created with the simulator introduced in \cref{sec:sim}.

Given that our goal is to show the capabilities of the framework and not the optimization itself, the simulator has been simplified to obtain a good number of samples in a reasonable amount of time.
It should be clear that such a rich database would correspondingly require more time when using a complex, thus more realistic simulation.

As usual in \gls{ml} when dealing with new datasets, the initial phase is devoted to the analysis of the gathered data.
A proper preprocessing, e.g., normalizing the inputs and removing the linearly correlated features, can boost the learning performance.
The scatter plot showing the relation between the inputs and the outputs is reported in \cref{fig:pairplot}.
Note that, even though the visual inspection of the data through different representations can help identify some hidden trends, its effectiveness is limited both by the high dimensionality of the problem and by the scarcity of available samples.
Therefore, in general, it is not possible to rely on this kind of data analysis for optimization.

The objective of the learning algorithm is to learn the underlying function mapping the input antenna configuration to the output network metrics, for example 
\begin{equation}
\begin{split}
  f:\mathbb{R}^n \to& \mathbb{R}^m\\
  \vb{x} \mapsto& \vb{y} 
\end{split}
\end{equation}
where $\vb{x}$ is the vector of the $n$ input antenna parameters, $f$ represents the simulator, computing the output network statistics from a given antenna configuration, and, finally, $\vb{y}$ is the vector of the $m$ considered network metrics.
Therefore, the learning algorithm (\emph{emulator}) learns an approximation $\hat{f}$ of the \emph{simulator}'s underlying function $f$, thus trying to mimic it.

\begin{figure*}[t]
    \centering
    \begin{subfigure}{0.49\textwidth}
        \centering
        \includegraphics[width=\textwidth]{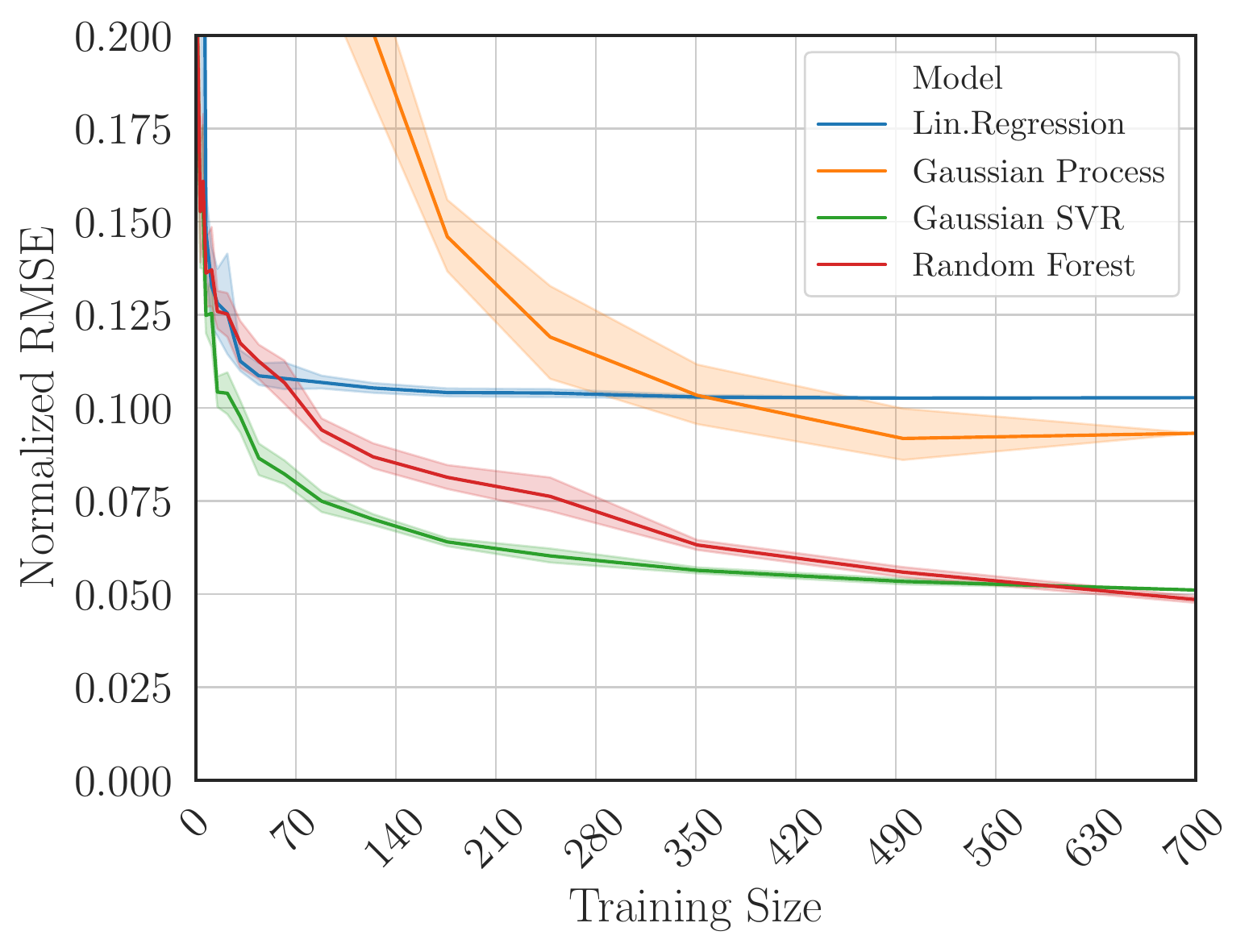}    
        \caption{$\sinr_5$}    
        \label{fig:incremental:a}
    \end{subfigure}
    \hfill
    \begin{subfigure}{0.49\textwidth}
        \centering
        \includegraphics[width=\textwidth]{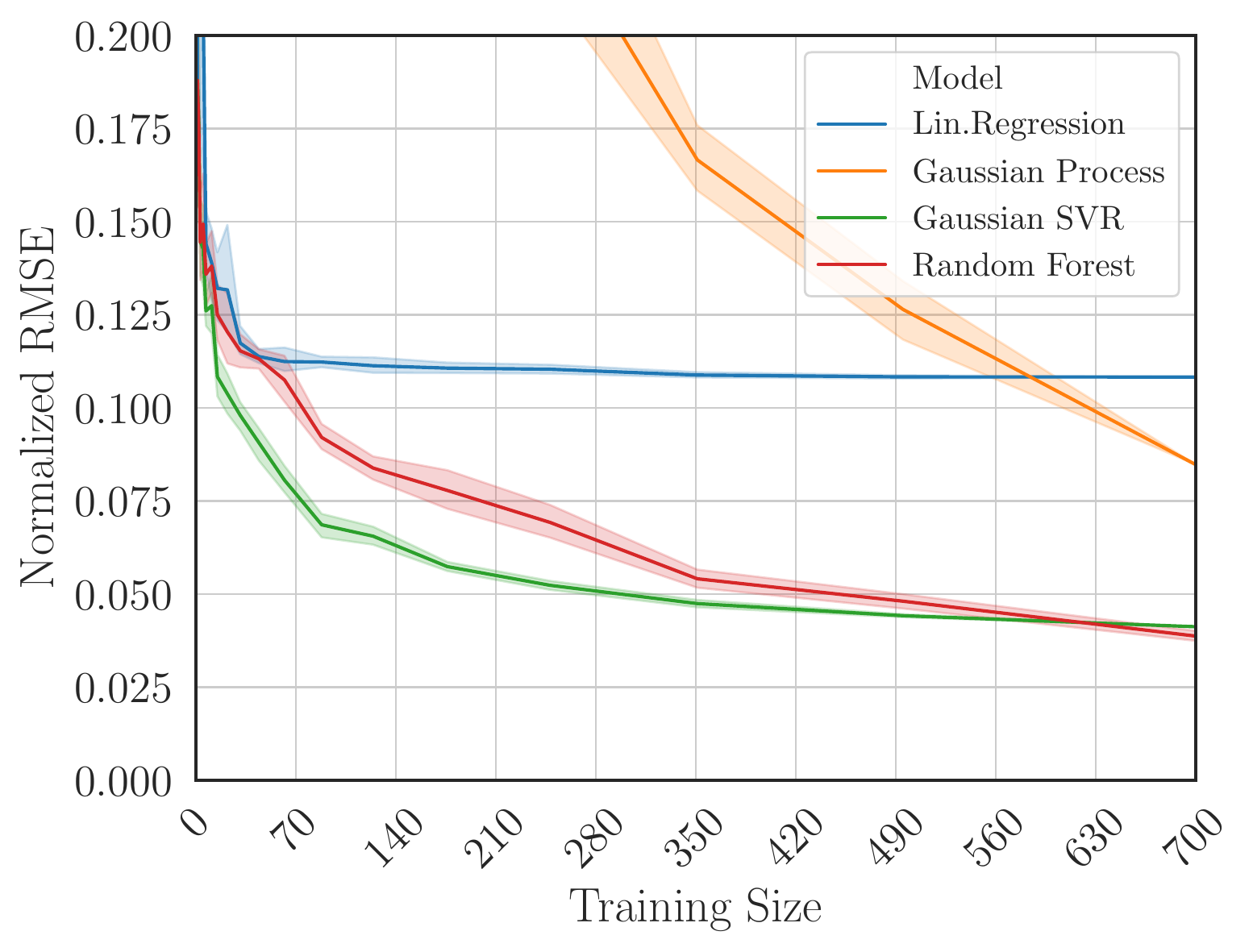}    
        \caption{\sinrmean}
        \label{fig:incremental:b}    
    \end{subfigure}
    \caption{Plots show the \gls{nrmse} as a function of the number of training samples.
    Multiple runs are performed, showing mean (line) and $95\%$ confidence interval (shadowed area) for each algorithm.}
    \label{fig:incremental}
\end{figure*}
Considering a scalar output $y$, the prediction or emulation error is then computed as the difference between the prediction of the emulator $\hat{y}$ and the corresponding simulator output $y$.
In order to assess this, we define the \gls{nrmse} as 
\begin{equation}
\label{nrmse_def}
\mathrm{nRMSE} = \sqrt{\frac{1}{N} \sum_{i=1}^{N}
    \qty( \frac{y_i-\hat{y}_i}
    {y_i} )^2 },
\end{equation}
where $N$ is the number of samples of the test set.
This parameter allows for a fair comparison among metrics on different scales, as the normalization yields a percentage of standard error with respect to the simulated value.
Note that \gls{sinr} values are first converted to linear units.

In this paper, results are validated using a test set of $300$ samples, that was proved to be large enough for this setup to obtain good testing accuracy.
In this work, the test set size is kept fixed, to allow a simplified presentation of the framework while guaranteeing a proper result validation.
Therefore, concerning question~\ref{item:nSamples}, only the training set size is taken into account.

Several learning techniques \cite{Bishop:2006:PRM:1162264} have been analyzed and tested.
However, only results for linear regression, \glspl{gpr}, random forests and \glspl{svr} are hereby reported.

\begin{itemize}
     \item \emph{Linear regression} is the most basic class of regression algorithms. Despite its simplicity, many versions and adaptations have been created, able to solve non-trivial problems. It is often considered as a baseline for more powerful algorithms.
     \item \emph{\glspl{gpr}} consider data as if it were sampled from a Gaussian stochastic process, trying to minimize the log-marginal-likelihood during the fit;
    \item \emph{Random forests} are ensembles of decision trees, that approximate stepwise the target function;
    \item \emph{\glspl{svr}} are derived from the \gls{svm} classification algorithm.
    Among all the typical kernels, the Gaussian one performed best and is used here.
\end{itemize}

One of the main advantages of linear regression is that, due to its simplicity, it is fast to train and easily interpretable, i.e., the analysis of the coefficients leads to some insights on the importance of the different inputs and their correlation.
On the other hand, random forests and \glspl{svr} are black-box algorithms, meaning that results are hardly interpretable.
Given their popularity, \glspl{nn} have been tested as well.
However, the lack of a large dataset has been found to be problematic for a stable convergence and they have thus been discarded from this study.

An effective way to address questions~\ref{item:1} and~\ref{item:nSamples} is reported in \cref{fig:incremental}, where the performance of the selected algorithms is evaluated for increasing training sizes.
We recall that increasing the number of training samples is always beneficial for learning, improving both emulation accuracy and stability.
However, it affects the dataset creation time, going against the purpose of the framework.
From the comparison of \cref{fig:incremental:a,fig:incremental:b}, it emerges that different emulation accuracies can be achieved for different metrics and that some learning algorithms predict a given metric better than others.
This suggests that the choice of the technique should be made specifically for each metric.
Eventually, this choice should also be made considering the number of available samples, as more complex algorithms, e.g., random forest, may outperform more basic ones when trained with enough samples.

Moreover, note that the performance of linear regression quickly saturate, while more complex algorithms achieve a lower error before converging.
Saturation is expected even with the most powerful algorithms since data obtained from the simulator is inherently noisy (e.g., the number of Monte Carlo simulations is never infinite, thus statistics are not perfect).
Instead, the reason why simpler algorithms tend to saturate earlier and with higher errors is because they are too simple to describe the inherent properties of the underlying function $f$.
This concept can be easily seen in \cref{fig:slice_fitting}, where we visually compare the emulator fit with the simulator samples.

As expected, the \gls{nrmse} decreases as the training size increases, but at different rates for different algorithms.
The trade-off between the number of samples and the emulation precision has to be taken into account when selecting the algorithm.
The achieved \gls{nrmse} can be quite low, namely about $3.2\%$ and $5.7\%$ for $\sinrmean$ and $\sinr_5$, respectively.
Finally, it can be observed that the two estimated metrics in \cref{fig:incremental} present different behaviors and performance for different outputs.
Furthermore, in general, we observed that it is not possible to have a universally valid list of best algorithms, as this is very much dependent on the simulator, the scenario, and even the considered metric.

As a basic approach, once the error achieves a target threshold, the emulator can be used for the optimization and the simulator can be stopped.
As an example, if our target is a $6\%$ error, then for $\sinr_5$ we would need $300$ training samples, while for $\sinrmean$ only $150$ samples would suffice, much fewer than the $700$ reported here.
Thus, in a realistic deployment, the number of required samples could be decided on the fly.

\subsection{Optimization}
The proposed framework is optimization agnostic, meaning that most standard numerical optimization techniques can be equally used.
Clearly, the learned representation is just an approximation of the real-world performance: while the simulator tries to reproduce reality via random experiments, the emulator tries to approximate the input-output relationship of the simulator via a black-box approach, adding a level of abstraction that further distances it from the real world.

\begin{figure}[t]
  \centering
  \includegraphics[width=1\columnwidth]{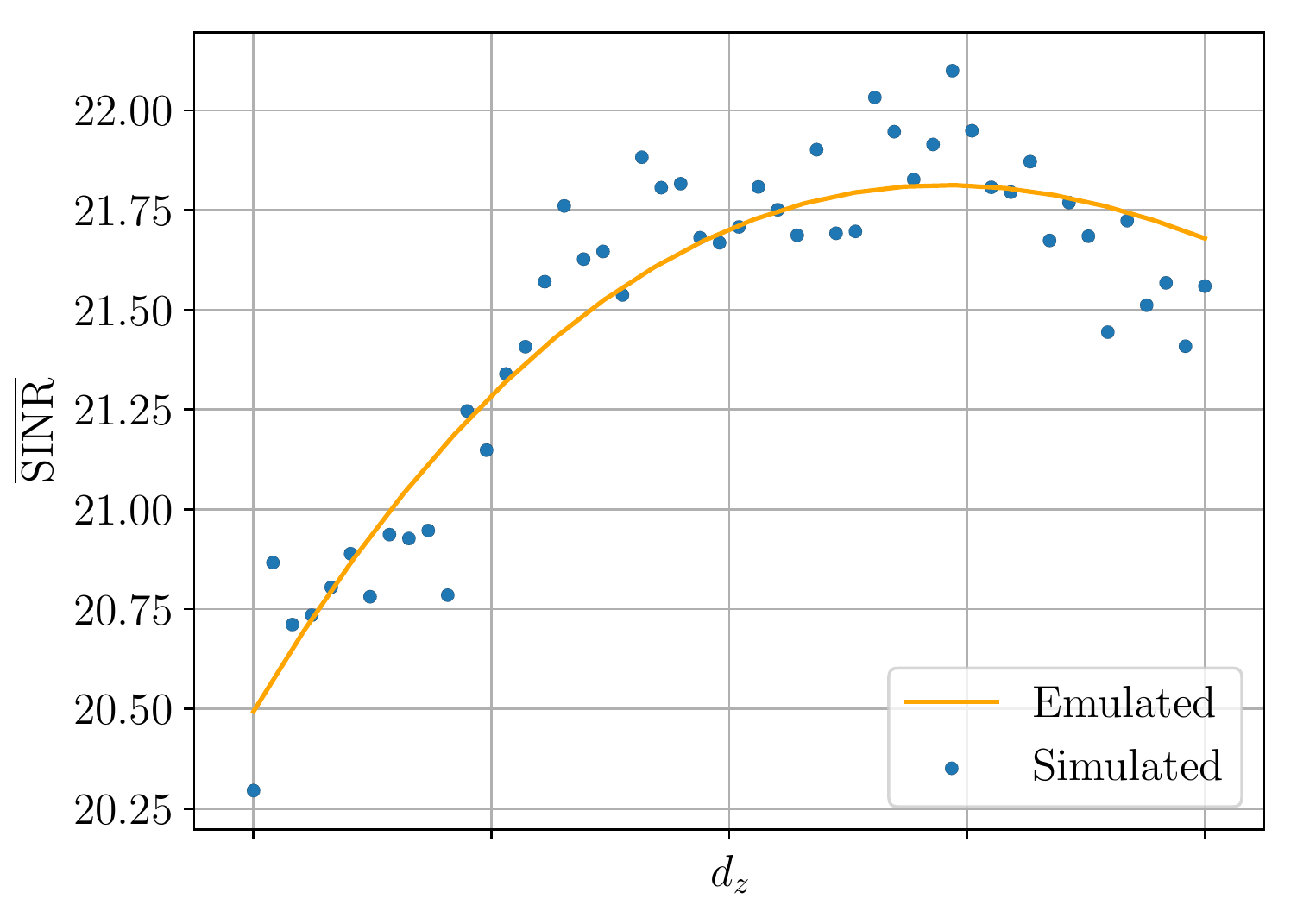}
  \caption{Representation of a one-dimensional plot obtained by fixing all the array parameters except one.
  The plot makes it possible to visually compare the emulator fit with the simulator samples.
  In this case, an $8\times8$ array was used with $d_y=0.5\lambda$ horizontal spacing, while the vertical spacing $d_z$ is varying.
  The emulator is still trained in all $4$ inputs simultaneously, justifying the suboptimal fit towards higher values of $d_z$.}
  \label{fig:slice_fitting}
\end{figure}
Since in general our models are not required to be differentiable, nor would we have an explicit derivative for most of them, gradient-based techniques are hardly usable.
Some of the inputs could also be categorical or discrete (e.g., the number of antennas in each dimension).
Furthermore, we are not posing any constraint on the convexity (or concavity) of the underlying function.
For these reasons, gradient-based optimization algorithms would not even be desirable.

On the other hand, since a global optimum is typically desired, gradient-free global optimization algorithms exist that satisfy all these requirements (e.g., genetic algorithms or simulated annealing).
Nowadays most scientific-oriented programming languages have optimization libraries, implementing several algorithms.
As briefly explained in \cref{sec:framework_description}, \cref{fig:slice_fitting} shows the noisiness of the training data.
Thus, finding the maximum values over the raw data might not be the best choice, while numerically finding a global maximum over a smooth model might be a better choice, provided that the model is not underfitting.
In the next section, we show the results obtained for the antenna optimization.

\section{Optimization Results}
\label{sec:optimization_results}
The optimization phase shows the significant advantages of this framework.
As previously stated, we remark that the proposed framework can be used for optimization in a wide set of scenarios, beyond that of cellular network design, used here as an example.
As the optimization is done jointly on all the input parameters, the hyperspace where it operates can be extremely vast and complex.
These features, along with the complexity of the search of the global maximum, require a very large number of evaluations.
The gain of the framework can then be measured comparing the number of entries necessary for the database creation with the number of function evaluations needed by the optimization.
This is because, due to the typical complexity of a simulator, the time required to obtain the database far exceeds that of the training and the optimization itself.
In terms of time costs, the training itself is negligible and, once trained, the predictions are instantaneous.

Another aspect to take into account is that, although significant, the database creation in our framework is an overhead that is needed only once, as it does not depend on the optimization goal.
The same emulator, providing almost instantaneous iterations, can be used with different optimization objectives, without requiring long simulation-based iterations.

For our example, given the data analysis initially done (partially shown in \cref{fig:pairplot}), we use as the objective function
\begin{equation}
\begin{split}
  &\text{maximize} \quad\quad \sinrmean\\
  &\text{s.t.} \quad\quad\quad \sinr_5 > 6\text{ dB}
\end{split}
\end{equation}
where the constraint on the worst \glspl{ut} (identified with $\sinr_5$) has been introduced in order to guarantee some degree of fairness and coverage to all the \glspl{ut} in the network.

\begin{figure}[t]
  \centering
  \includegraphics[width=1\columnwidth]{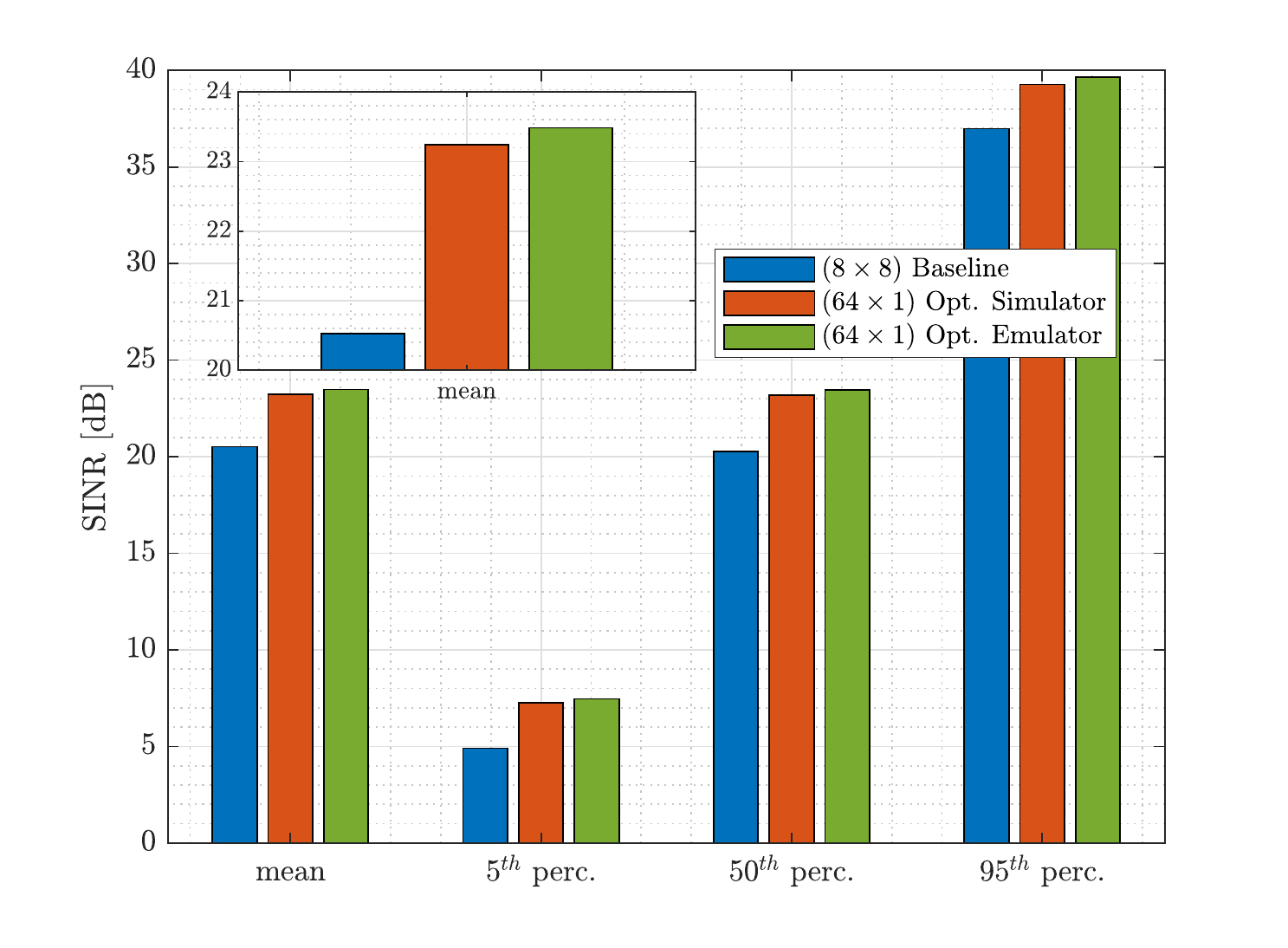}
  \caption{Comparison among the network performance obtained with the baseline configuration (blue bar), with the optimal configuration identified using the simulator samples (orange bar) and using the emulator (green bar).}
  \label{fig:opt_results}
\end{figure}
\begin{table}[t]
\centering
\caption{Numerical results shown in \cref{fig:opt_results}.}
\label{tab:opt_results}
\begin{tabular}{lcccc}
\toprule
            & \sinrmean & $\sinr_5$ & $\sinr_{50}$ & $\sinr_{95}$ \\
\midrule
Baseline    & $20.52$           & $4.91$    & $20.26$      & $36.99$      \\
Opt. Simulator & $23.24$           & $7.25$    & $23.18$      & $39.27$      \\
Opt. Emulator  & $23.49$           & $7.47$    & $23.45$      & $39.64$     \\
\bottomrule
\end{tabular}
\end{table}
The optimization results obtained within the scenario described in the previous section are presented in \cref{tab:opt_results,fig:opt_results}.
Results show that in the proposed scenario, a baseline setup consisting of $8\times 8$ arrays with $\lambda/2$ spacing in both directions performs significantly worse than the optimized ones.
The other two configurations represent the optimum obtained over the collected dataset (\emph{Opt. Simulator}, made of $1,000$ randomly sampled points in the four-dimensional space described in \cref{sec:framework_description}) and the global optimum obtained using our framework (\emph{Opt. Emulator}).
They both identified a $64\times 1$ configuration (vertical \gls{ula}), but respectively with $0.825\lambda$ and $0.734\lambda$ spacing.
Results show a $\sim 3$~dB improvement over the trivial baseline.
Although in this case the results are really close (both inputs and outputs), two facts are important: firstly, we discussed in \cref{sec:framework_description} that significantly fewer than $1,000$ samples would have been enough, a far lower number than required by a brute force optimization; secondly, as more inputs are considered, the input space will not be sampled enough to find a good setup, making emulation even more important.

Having computed $1,000$ samples while the optimization required more than $12,000$ function evaluations, we obtain a speedup factor of $12\times$ with respect to brute force evaluation.
A key advantage of our approach is the possibility of changing the objective functions of the optimizer, which would be easily and quickly done with the emulator, without having to retrain it.

\section{Conclusions and Future works}
\label{sec:conclusions}
An innovative framework has been presented that makes the joint optimization of multiple parameters a reality, needing just a fraction of the time that is currently required when directly employing a simulator.
As simulators are generally computationally complex and time-consuming, the key idea is to bypass them using a fast emulator, obtained through \gls{ml} techniques.
After a long, initial database creation, any objective function can be optimized in a matter of minutes or even seconds.
The effectiveness of this methodology has been proved using a network simulator.
Network simulators require a long time to compute the network metrics for specific antenna configurations, thus representing the perfect testbed for our framework.

Future works call, in the first place, for further studies on how to reduce the number of required training samples, in order to further reduce the dataset creation overhead.
Moreover, a second aspect would be to increase the accuracy of the emulators, possibly resorting to more complex \gls{ml} techniques.
Finally, the range of applicability of the framework, concerning both the complexity of the involved simulator and the number of parameters to be optimized, is left for future studies.

\section*{Acknowledgments}
Mattia Lecci's activity was financially supported by \emph{Fondazione Cassa di Risparmio Padova e Rovigo} under the grant ``Dottorati di Ricerca 2018''.

\bibliographystyle{IEEEtran}
\bibliography{bibl}

\end{document}